\documentclass[12pt]{article}
\usepackage{graphicx}
\usepackage{amsmath}
\usepackage{hhline}
\usepackage{amssymb}
\usepackage{times}
\usepackage{cite}
\usepackage{array}
\usepackage{multirow}

\usepackage{color}
\definecolor{rltred}{rgb}{0.75,0,0}
\definecolor{rltgreen}{rgb}{0,0.5,0}
\definecolor{rltblue}{rgb}{0,0,0.75}

\newif\ifpdf
\ifx\pdfoutput\undefined
    \pdffalse          
\else
    \pdfoutput=1       
    \pdftrue
\fi

\ifpdf
\usepackage{thumbpdf}
\usepackage[pdftex,
        colorlinks=true,
        urlcolor=rltblue,       
        filecolor=rltgreen,     
        linkcolor=rltred,       
        pdftitle={Example File for pdflatex},
        pdfauthor={H1},
        pdfsubject={Template file},
        pdfkeywords={High-Energy Physics, Particle Physics},
        pdfpagemode=None,
        bookmarksopen=true]{hyperref}
 
\fi

\newlength{\dinwidth}
\newlength{\dinmargin}
\begin{document}

\newcommand{\pom}{{I\!\!P}}
\newcommand{\reg}{{I\!\!R}}
\newcommand{\slowpi}{\pi_{\mathit{slow}}}
\newcommand{\fiidiii}{F_2^{D(3)}}
\newcommand{\fiidiiiarg}{\fiidiii\,(\beta,\,Q^2,\,x)}
\newcommand{\n}{1.19\pm 0.06 (stat.) \pm0.07 (syst.)}
\newcommand{\nz}{1.30\pm 0.08 (stat.)^{+0.08}_{-0.14} (syst.)}
\newcommand{\fiidiiiful}{F_2^{D(4)}\,(\beta,\,Q^2,\,x,\,t)}
\newcommand{\fiipom}{\tilde F_2^D}
\newcommand{\ALPHA}{1.10\pm0.03 (stat.) \pm0.04 (syst.)}
\newcommand{\ALPHAZ}{1.15\pm0.04 (stat.)^{+0.04}_{-0.07} (syst.)}
\newcommand{\fiipomarg}{\fiipom\,(\beta,\,Q^2)}
\newcommand{\pomflux}{f_{\pom / p}}
\newcommand{\nxpom}{1.19\pm 0.06 (stat.) \pm0.07 (syst.)}
\newcommand {\gapprox}
   {\raisebox{-0.7ex}{$\stackrel {\textstyle>}{\sim}$}}
\newcommand {\lapprox}
   {\raisebox{-0.7ex}{$\stackrel {\textstyle<}{\sim}$}}
\def\gsim{\,\lower.25ex\hbox{$\scriptstyle\sim$}\kern-1.30ex%
\raise 0.55ex\hbox{$\scriptstyle >$}\,}
\def\lsim{\,\lower.25ex\hbox{$\scriptstyle\sim$}\kern-1.30ex%
\raise 0.55ex\hbox{$\scriptstyle <$}\,}
\newcommand{\pomfluxarg}{f_{\pom / p}\,(x_\pom)}
\newcommand{\dsf}{\mbox{$F_2^{D(3)}$}}
\newcommand{\dsfva}{\mbox{$F_2^{D(3)}(\beta,Q^2,x_{I\!\!P})$}}
\newcommand{\dsfvb}{\mbox{$F_2^{D(3)}(\beta,Q^2,x)$}}
\newcommand{\dsfpom}{$F_2^{I\!\!P}$}
\newcommand{\gap}{\stackrel{>}{\sim}}
\newcommand{\lap}{\stackrel{<}{\sim}}
\newcommand{\fem}{$F_2^{em}$}
\newcommand{\tsnmp}{$\tilde{\sigma}_{NC}(e^{\mp})$}
\newcommand{\tsnm}{$\tilde{\sigma}_{NC}(e^-)$}
\newcommand{\tsnp}{$\tilde{\sigma}_{NC}(e^+)$}
\newcommand{\st}{$\star$}
\newcommand{\sst}{$\star \star$}
\newcommand{\ssst}{$\star \star \star$}
\newcommand{\sssst}{$\star \star \star \star$}
\newcommand{\tw}{\theta_W}
\newcommand{\sw}{\sin{\theta_W}}
\newcommand{\cw}{\cos{\theta_W}}
\newcommand{\sww}{\sin^2{\theta_W}}
\newcommand{\cww}{\cos^2{\theta_W}}
\newcommand{\trm}{m_{\perp}}
\newcommand{\trp}{p_{\perp}}
\newcommand{\trmm}{m_{\perp}^2}
\newcommand{\trpp}{p_{\perp}^2}
\newcommand{\alp}{\alpha_s}

\newcommand{\alps}{\alpha_s}
\newcommand{\sqrts}{$\sqrt{s}$}
\newcommand{\LO}{$O(\alpha_s^0)$}
\newcommand{\Oa}{$O(\alpha_s)$}
\newcommand{\Oaa}{$O(\alpha_s^2)$}
\newcommand{\PT}{p_{\perp}}
\newcommand{\JPSI}{J/\psi}
\newcommand{\sh}{\hat{s}}
\newcommand{\uh}{\hat{u}}
\newcommand{\MP}{m_{J/\psi}}
\newcommand{\PO}{I\!\!P}
\newcommand{\xbj}{x}
\newcommand{\xpom}{x_{\PO}}
\newcommand{\ttbs}{\char'134}
\newcommand{\xpomlo}{3\times10^{-4}}  
\newcommand{\xpomup}{0.05}  
\newcommand{\dgr}{^\circ}
\newcommand{\pbarnt}{\,\mbox{{\rm pb$^{-1}$}}}
\newcommand{\gev}{\,\mbox{GeV}}
\newcommand{\mev}{\,\mbox{MeV}}
\newcommand{\WBoson}{\mbox{$W$}}
\newcommand{\fbarn}{\,\mbox{{\rm fb}}}
\newcommand{\fbarnt}{\,\mbox{{\rm fb$^{-1}$}}}

%
%
\newcommand{\qsq}{\ensuremath{Q^2} }
\newcommand{\gevsq}{\ensuremath{\mathrm{GeV}^2} }
\newcommand{\et}{\ensuremath{E_t^*} }
\newcommand{\rap}{\ensuremath{\eta^*} }
\newcommand{\gp}{\ensuremath{\gamma^*}p }
\newcommand{\dsiget}{\ensuremath{{\rm d}\sigma_{ep}/{\rm d}E_t^*} }
\newcommand{\dsigrap}{\ensuremath{{\rm d}\sigma_{ep}/{\rm d}\eta^*} }
\def\Journal#1#2#3#4{{#1} {\bf #2} (#3) #4}
\def\NCA{\em Nuovo Cimento}
\def\NIM{\em Nucl. Instrum. Methods}
\def\NIMA{{\em Nucl. Instrum. Methods} {\bf A}}
\def\NPB{{\em Nucl. Phys.}   {\bf B}}
\def\PLB{{\em Phys. Lett.}   {\bf B}}
\def\PRL{\em Phys. Rev. Lett.}
\def\PRD{{\em Phys. Rev.}    {\bf D}}
\def\ZPC{{\em Z. Phys.}      {\bf C}}
\def\EJC{{\em Eur. Phys. J.} {\bf C}}
\def\CPC{\em Comp. Phys. Commun.}

\begin{titlepage}

\noindent
\begin{flushleft}
DESY 06-020\hfill ISSN 0418-9833\\
March 2006
\end{flushleft}

\vspace{2cm}

\begin{center}
\begin{Large}

{\bf Photoproduction of Dijets with High Transverse Momenta at HERA}

\vspace{2cm}

H1 Collaboration

\end{Large}
\end{center}

\vspace{2cm}

\begin{abstract}
\noindent
Differential dijet cross sections are measured in photoproduction in the region 
of photon virtualities $Q^{2} < 1 \gev^{2} $ with the H1 detector at the HERA 
$ep$ collider using an integrated luminosity of $66.6$~$\rm{pb}^{-1}$.
Jets are defined with the inclusive $k_{\rm{\perp}}$ algorithm and a 
minimum transverse momentum of the leading jet of $25\gev$ is required.
Dijet cross sections are measured in direct and resolved photon enhanced regions separately.
Longitudinal proton momentum fractions up to $0.7$ are reached.
The data compare well with predictions from Monte Carlo event generators based on 
leading order QCD and parton showers and with next-to-leading order QCD calculations 
corrected for hadronisation effects.

\end{abstract}

\vspace{1.5cm}

\begin{center}
Submitted to Phys. Lett. B
\end{center}

\end{titlepage}

%
%
%
\begin{flushleft}

A.~Aktas$^{9}$,                
V.~Andreev$^{25}$,             
T.~Anthonis$^{3}$,             
B.~Antunovic$^{26}$,           
S.~Aplin$^{9}$,                
A.~Asmone$^{33}$,              
A.~Astvatsatourov$^{3}$,       
A.~Babaev$^{24}$,              
S.~Backovic$^{30}$,            
J.~B\"ahr$^{38}$
% ...

\bigskip{\it
% ...
$ ^{3}$ Inter-University Institute for High Energies ULB-VUB, Brussels;
          Universiteit Antwerpen, Antwerpen; Belgium$^{ c}$ \\
% ...
$ ^{9}$ DESY, Hamburg, Germany \\
% ...
$ ^{24}$ Institute for Theoretical and Experimental Physics,
          Moscow, Russia$^{ l}$ \\
 $ ^{25}$ Lebedev Physical Institute, Moscow, Russia$^{ e}$ \\
 $ ^{26}$ Max-Planck-Institut f\"ur Physik, M\"unchen, Germany \\
% ...
$ ^{30}$ Faculty of Science, University of Montenegro,
          Podgorica, Serbia and Montenegro$^{ e}$ \\
% ...
$ ^{33}$ Dipartimento di Fisica Universit\`a di Roma Tre
          and INFN Roma~3, Roma, Italy \\
% ...
$ ^{38}$ DESY, Zeuthen, Germany \\
% ...
$ ^c$ Supported by FNRS-FWO-Vlaanderen, IISN-IIKW and IWT
      and  by Interuniversity
Attraction Poles Programme,
      Belgian Science Policy \\
% ...
$ ^e$ Supported by the Deutsche Forschungsgemeinschaft \\
% ...
$ ^l$ Partially Supported by Russian Foundation
      for Basic Research,  grants  03-02-17291
      and  04-02-16445 \\
}
\end{flushleft}

\newpage

\section{Introduction}
\noindent
At HERA the largest cross section is due to photoproduction, where the beam lepton interacts with the proton 
via the exchange of  a photon at small virtualities $Q^2 \approx 0$. The photoproduction of dijets with high 
transverse momenta can be calculated within perturbative Quantum Chromodynamics (pQCD) where the transverse 
momentum of jets provides the hard scale.

Two contributions to the jet cross section can be distinguished: {\it direct processes} in which the photon itself 
enters the hard subprocess and {\it resolved processes} in which the photon fluctuates into partons of which one 
participates in the hard scatter. The hadronic structure of the proton and photon are described by their respective 
parton density functions (PDFs).

Measurements of the parton densities of the photon and proton have been performed in several processes 
in $e^+e^-$, $ep$ and $p \bar{p}$ collisions. 
The quark densities in the photon have been determined at $e^+e^-$ colliders. 
The parton densities of the proton are mainly determined from deep inelastic scattering (DIS) experiments. 
Drell-Yan and $p \bar{p}$ jet data provide constraints on the gluon density at high longitudinal proton momentum fraction ($x_p$). 
Previous dijet data in photoproduction~\cite{zeusjets}, as well as in electroproduction,
are shown to constrain the gluon density in the medium $x_p$ 
region~\cite{zeusfits,wobisch}. Compared to $e^+e^-$
data the photoproduction of jets reaches higher scales and is directly sensitive to the 
gluon density in the photon.

To test predictions of perturbative calculations and current PDF parametrisations this paper investigates 
dijet production at very small $Q^2$ in positron proton interactions using the H1 detector at HERA. The 
transverse momentum ($E_{t}$) of the leading jet ranges between $25$ and $80\gev$. The range of the photon 
momentum fraction carried by the parton participating in the hard interaction is $0.1 < x_\gamma < 1.0$. The 
proton momentum fraction carried by the interacting parton from the proton side is in the range of $0.05 < x_p < 0.7$. 

This paper, compared to a previous publication~\cite{h1dijet:2002}, presents new measurements with increased statistical 
precision and an improved understanding of the systematic uncertainties. In addition, new measurements are made which 
examine cross sections with different jet topologies. The dijet cross sections are compared with Monte Carlo simulations 
based on leading order (LO) QCD and parton showers and with next-to-leading order (NLO) pQCD calculations with hadronisation corrections.

\section{H1 Detector}
\noindent
The H1 detector is described in detail elsewhere~\cite{h1det}. 
The detector elements important for this analysis are described below. 

The liquid argon (LAr) calorimeter covers a range in polar angle\footnote{H1 uses a right-handed coordinate 
system with the $z$-axis along the direction of the outgoing proton beam. The polar angle $\theta$ is defined 
with respect to the $z$-axis.} of $4^{\circ}~<~\theta~<~153^{\circ}$. The angular region 
$153^{\circ} < \theta < 177^{\circ}$ is covered by the SpaCal (a lead scintillating fibre Spaghetti Calorimeter). 
The central tracking detector consists of two concentric drift chambers supplemented by two $z$-drift chambers
and has an angular coverage of 
$25^{\circ} < \theta < 155^{\circ}$. These detectors are immersed in a $1.15~\rm{T}$ magnetic field.

The LAr and SpaCal calorimeters are used to trigger events and to reject non photoproduction events which have 
an identified scattered positron. Together with the central tracking chambers they provide a measurement of the 
hadronic final state energies from which jets are reconstructed. The central tracking chambers are also used to 
reconstruct the event vertex.
 
The luminosity determination is based on the measurement of the Bethe-Heitler process ($ ep\rightarrow ep\gamma $), 
where the photon is detected in a calorimeter located downstream of the interaction point in the positron beam direction.

\section{Event Selection}
\noindent
The data used in this analysis were taken in the years 1999-2000 where positrons of energy $27.6 \gev $ were collided 
with protons of $920 \gev $, yielding a centre-of-mass energy of $318 \gev$. This sample corresponds to a total 
integrated luminosity of $66.6$~$\pbarnt$.

Events were triggered by requiring a combination of sub-triggers utilising different energy thresholds in the LAr 
calorimeters with additional vertex and timing conditions. The trigger efficiency is above $98\%$ for the event 
selection used in this analysis.

The event vertex is required to be reconstructed within $\pm 35~\rm{cm}$ in $z$ of the nominal interaction point. This 
ensures that the event can be properly reconstructed and helps to remove proton beam-gas background events. Several 
topological background finder algorithms are used to remove cosmic muon events. Events with a large missing transverse 
momentum of more than $20 \gev$ are rejected, reducing charged current and any remaining non-$ep$ background to 
below the $1\%$ level. 

Photoproduction events are selected by demanding that there be no scattered
positron candidate in the LAr or SpaCal calorimeter, restricting the negative
four-momentum transfer squared $Q^2$ to be below $1 \gev^2$.
The main source of background comes from neutral current (NC) DIS events in which the scattered positron 
is misidentified as part of the hadronic final state. 
These events are suppressed by requiring the inelasticity $y$ to be less than $0.9$, where $y$ 
is reconstructed from the hadronic final state\footnote{The inclusion of the scattered positron 
into the hadronic final state causes $y$ to be reconstructed at values close to one.}. 
The phase space is further restricted to $y > 0.1$. Additional restrictions based on the topology of the jet 
showers~\cite{ingo} are applied that help to reduce the overall DIS background to below $2\%$. The remaining 
DIS background is subtracted statistically based on predictions from Monte Carlo simulations.

Jets are reconstructed in the laboratory frame using the inclusive $k_{\rm{\perp}}$ algorithm~\cite{Catani:1993hr}. 
The $p_{\rm{t}}$-weighted recombination scheme is used in which the jets are considered massless and the 
separation parameter is set to 1. 
The jets are required to be contained in the LAr calorimeter by the restriction that 
$-0.5 < \eta_{\rm{jet}} < 2.75 $, where the pseudo-rapidity is given by $\eta = -\ln{\tan{\theta/2}}$. 
Only the two highest $E_{\rm{t}}$ jets in the chosen $\eta$ range are considered. 
Asymmetric cuts on the jets $E_{\rm{t}}$ are applied to avoid regions of phase space where 
the existing NLO QCD calculations suffer from an incomplete cancellation of infrared singularities. 
The leading jet is required to have $E_{\rm{t,max}}>25 \gev$ and the other jet $E_{\rm{t,2nd}}>15 \gev$. 
The total number of selected events within the phase space summarised in table~\ref{tab:phasespace} is about $14,000$.

\begin{table}[htdp]
\begin{center}
\begin{tabular}{c}
$Q^{2} < 1 \gev^{2}$ \\
\vspace{1mm}
$0.1 < y < 0.9$\\
\vspace{1mm}
$E_{\rm{t,max}} > 25 \gev$\\
\vspace{1mm}
$E_{\rm{t,2nd}} > 15 \gev$\\
\vspace{1mm}
$-0.5 < \eta_{\rm{jet}} < 2.75$\\
\end{tabular}
\end{center}
\caption{Definition of the phase space of the dijet cross section measurements.}
\label{tab:phasespace}
\end{table}

\section{Jet Observables}
\noindent
This analysis studies the dijet cross section as a function of the two observables $x_{\gamma}$ and $x_{p}$ 
and as a function of the angle of the dijets in their centre-of-mass system, $|\cos \theta^{*}|$. These 
variables are reconstructed as follows:

\begin{equation}
  \label{eq:recxgamma}
  x_\gamma  =  \frac{1}{2yE_e} \cdot \sum_{i=1}^2 E_{\rm{t,i}} \cdot e^{-\eta_i}
\end{equation}

\begin{equation}
  \label{eq:recxproton}
  x_p  =  \frac{1}{2E_p} \cdot \sum_{i=1}^2 E_{\rm{t,i}} \cdot e^{+\eta_i}
\end{equation}

\begin{equation}
  \label{eq:costheta}
  |\cos \theta^{*}| = | \tanh(\eta_{1} - \eta_{2})/2 | \, .
\end{equation}

\noindent
Here $E_e$ and $E_p$ are the energies of the positron and proton beam, respectively. 
$E_{\rm{t,1}}$ and $E_{\rm{t,2}}$ are the transverse energies of the two jets and $\eta_{\rm{1}}$ and 
$\eta_{\rm{2}}$ their pseudorapidities. In the leading order picture $x_\gamma$ and $x_p$ 
represent, respectively, the 
longitudinal photon and proton momentum fractions entering the hard interaction.

\section{QCD Models}
\noindent
The PYTHIA~\cite{Sjostrand} Monte Carlo program contains Born level QCD matrix elements of direct and resolved 
hard processes. Higher order QCD radiation is represented by parton showers in the leading logarithm approximation. 
PYTHIA uses the Lund string model for hadronisation. Here version 6.1 of PYTHIA is used with the 
leading order parametrisation CTEQ5L~\cite{CTEQ5} for the proton PDFs and GRV-LO~\cite{GRV} for the photon PDFs. 
The PYTHIA predictions need to be scaled up by a factor of $1.2$ to 
describe the dijet data, this factor accounting for missing
higher orders in the PYTHIA calculation.

The HERWIG~\cite{Marchesini} Monte Carlo, which uses the cluster model for hadronisation, is found to produce 
similar results to PYTHIA, but a scale factor of 1.55 is required to reproduce the total dijet cross section.

Parton level NLO QCD dijet cross sections are obtained using a program~\cite{Frixione} based on the subtraction 
method~\cite{Kunszt:1992tn} for the cancellation of infrared singularities. In the calculation of the NLO cross 
sections a two-loop $\alpha_s$ is taken and the parametrisation CTEQ6M~\cite{cteq6} is chosen. 
Using instead the MRST2001~\cite{mrst2001} PDFs similar results are found. 
The uncertainty of the NLO QCD predictions due to the choice of the proton PDFs is calculated from 
the 40 eigenvectors of the CTEQ6M PDFs. It varies from $4\%$ at low $x_p$ to $20\%$ at high $x_p$. 
For the photon PDFs the GRV-HO~\cite{grv} parametrisation is used. Using instead the AFG-HO~\cite{agf}
photon PDFs, differences of the order of $20\%$ in the resolved enhanced region and of $10\%$ in the direct 
enhanced region are seen~\cite{ingo}.

The renormalisation scale $\mu_r$ and the factorisation scale $\mu_f$ are set to the sum of the transverse 
momenta of the outgoing partons divided by two, on an event-by-event basis. The effect of the choice of scale 
was studied by varying the common scale $\mu = \mu_r = \mu_f$ by a factor two up and one half down. The uncertainty 
on the NLO QCD predictions arising from this procedure is found to vary between a few percent and almost $\pm 30\%$. 
The uncertainty from the PDFs is in general much smaller than the error from the scale uncertainty, except at 
large $x_{p}$ where it grows to be about twice as big.

The NLO QCD predictions are compared to the data after a correction for hadronisation effects. The correction 
$\delta_{\rm{had}}$ is determined from the Monte Carlo models and varies between $1\%$ and $6\%$. It is defined 
as the ratio of the cross sections calculated with jets reconstructed from hadrons to those from partons 
(after the parton shower). HERWIG and PYTHIA are used to calculate a mean correction factor applied to the NLO 
QCD predictions. Its uncertainty is taken as half the difference between the HERWIG and PYTHIA results. 
This uncertainty ($3$ - $6\%$) is in general smaller than the dominant theory uncertainty which depending on 
phase space is given by the scale or PDF uncertainty.

\section{Data Correction}
\label{sec:datacorr}
\noindent
The data are corrected for detector effects (resolution and efficiencies) using Monte Carlo
event samples.
The correction factors are calculated from the ratio of the cross sections with jets reconstructed from 
hadrons (hadron level) and from detector objects (detector level). The correction is applied bin-by-bin. 
The bin sizes used in the cross section measurements are matched to the resolution and generally result 
in high acceptance and 
purity\footnote{The acceptance (purity) is defined as the ratio of the number of events generated 
in a bin which are reconstructed in that bin to the total number of events generated (reconstructed) 
in that bin.}, typically above $60\%$, with a minimum requirement of $30\%$. 
The Monte Carlo events are reweighted to take into account the imperfect description of the observed 
$y$ and $|\cos\theta^{*}|$ distributions. Both HERWIG and PYTHIA produce similar correction factors 
and a mean correction factor is used. The uncertainty  in the correction factor is taken as half the 
difference between HERWIG and PYTHIA. Half the difference between the reweighted and unweighted results 
is taken as an additional uncertainty.

\section{Systematic Uncertainties}
\noindent
For the jet cross sections the following sources of correlated and uncorrelated systematic errors are considered.
\begin{itemize}
\item The LAr hadronic energy scale is known to within $1.5\%$. 
It is estimated from the $p_t$ balance of the scattered positron with the hadronic final state 
in DIS and from the $p_t$ balance of dijet events in photoproduction in the $p_t$ range of this 
analysis. The resulting correlated uncertainties on the cross section are typically 
$7\%$ at low $x_p$ and $15\%$ at high $x_p$.
\item The SpaCal hadronic energy scale is known to better than $8\%$, resulting in correlated uncertainties of typically $1\%$ to $2\%$.
\item The total uncertainty in the data correction factor (see section~\ref{sec:datacorr}) results in a typical error of $2\%$ to $7\%$ and is considered as uncorrelated.
\item The trigger efficiency uncertainty results in an uncorrelated error of $2\%$.
\item The subtraction of the DIS background leads to an uncorrelated error of less than $1\%$.
\item The uncertainty in the luminosity measurement leads to an overall normalisation error of $1.5\%$.
\end{itemize}

\section{Results}
\begin{figure}[e]
  \begin{center}
    \includegraphics[width=\textwidth]{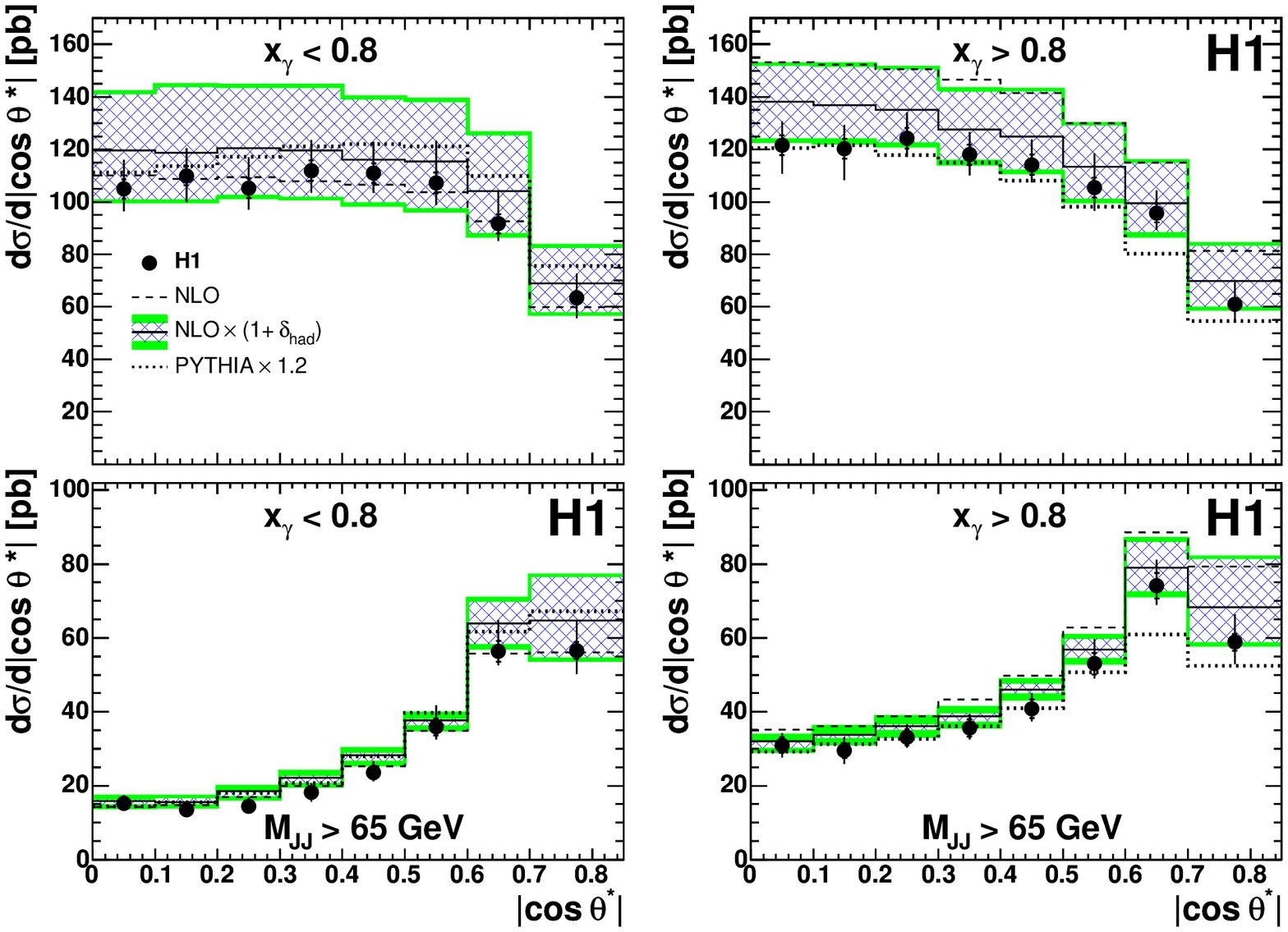}
    \caption{Bin averaged cross sections as a function of $|\cos\theta^{*}|$ for data (points), 
      NLO QCD calculations with (solid line) and without (dashed line) hadronisation corrections 
      $\delta_{\rm{had}}$ and for the PYTHIA Monte Carlo predictions (dotted line) scaled by a factor of 
      $1.2$. The inner bars indicate the statistical uncertainty and the outer error bars show the 
      statistical and systematic errors added in quadrature. The inner (hatched) band of the 
      NLO$\times (1+\delta_{\rm{had}})$ result is the scale uncertainty, the outer (shaded) band is the total 
      uncertainty. The cross sections are shown for two regions in $x_\gamma$ enhancing the resolved (left) 
      or direct (right) photon contribution,
      with and without an additional cut applied on the invariant dijet
      mass ($M_{\rm{JJ}}$).}
    \label{fig:costheta} 
  \end{center}
\end{figure}

\noindent
The dijet cross section as a function of $|\cos\theta^{*}|$ is shown in 
figure~\ref{fig:costheta} and listed in table~\ref{table:costheta}. This distribution 
is sensitive to the dynamics of the hard interaction. The measurement is presented for 
the direct ($x_\gamma > 0.8$) and resolved ($x_\gamma < 0.8$) enriched samples separately. 
The cross section shows no enhancement in the region of large $|\cos \theta^{*}|$ because 
the cuts on the jet transverse momenta suppress the phase space in this region. This phase 
space suppression is less prominent for large energies in the centre-of-mass of the hard 
subprocess. Requiring in addition that the dijet mass $M_{JJ}$ be above $65 \gev$, the shape 
of the measured cross section is changed towards that expected from the QCD matrix elements. 
The cross section in the resolved sample rises more rapidly with $|\cos \theta^{*}|$ than 
that in the direct sample due to the dominating gluon propagator in resolved 
processes\cite{h1dijet:2002, zeusjets}.

\begin{figure}[e]
  \begin{center}
    \includegraphics[width=\textwidth]{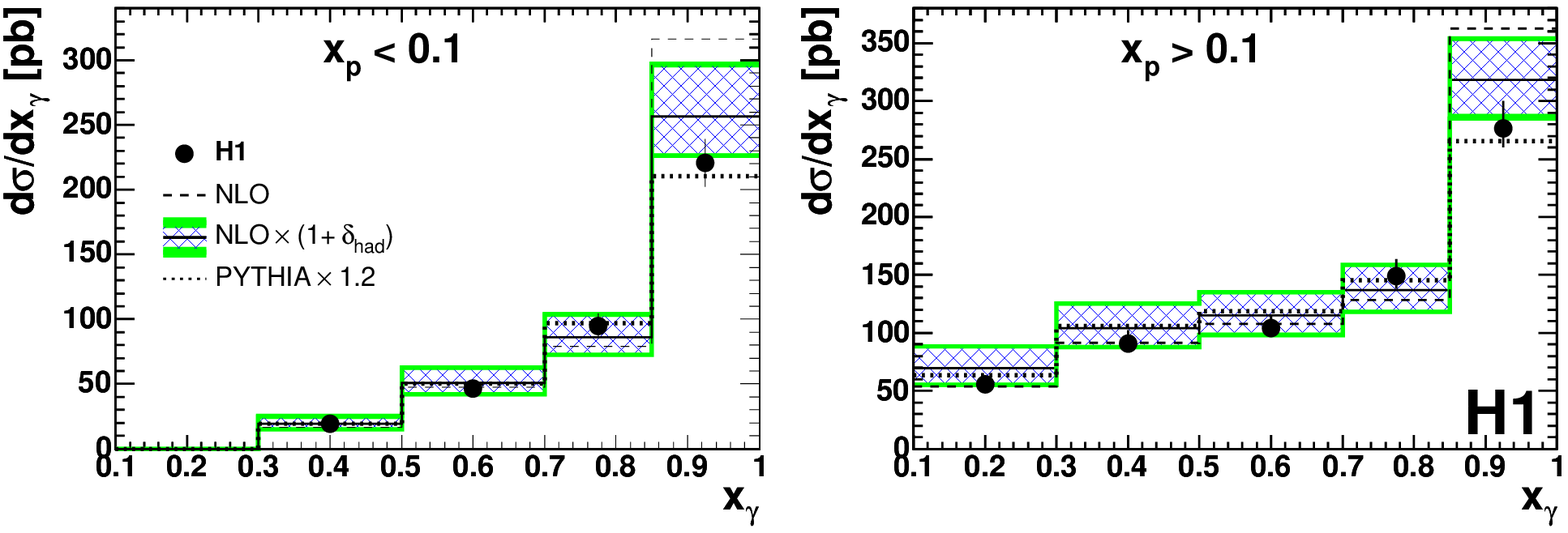}
    \caption{Bin averaged cross sections as a function of $x_\gamma$ for data (points), NLO QCD calculations 
       with (solid line) and without (dashed line) hadronisation corrections $\delta_{\rm{had}}$ and for the PYTHIA 
       Monte Carlo predictions (dotted line) scaled by a factor of $1.2$. The inner bars indicate the statistical 
       uncertainty and the outer error bars show the statistical and systematic errors added in quadrature. The 
       inner (hatched) band of the NLO$\times (1+\delta_{\rm{had}})$ result is the scale uncertainty, the outer 
       (shaded) band is the total uncertainty. The cross sections are shown separately for two regions in $x_p$.}
    \label{fig:xgamma} 
  \end{center}
\end{figure}

Figure~\ref{fig:xgamma} (table~\ref{table:xgamma}) shows the cross section as a function of 
$x_\gamma$ in two regions of $x_p$. For $x_p < 0.1$ the fraction of events induced by gluons 
from the proton side is estimated to be about $70\%$. It decreases to $15\%$ at the highest 
$x_p$ reached in this analysis. Thus the two regions roughly distinguish between photon-gluon 
fusion ($x_p < 0.1$) and photon-quark scattering ($x_p > 0.1$). Over the entire range in  
$x_\gamma$ and in both $x_p$ regions the NLO QCD predictions agree with the data within uncertainties. 
The leading order Monte Carlo predictions also describe the data.

\begin{figure}[e]
  \begin{center}
     \includegraphics[width=\textwidth]{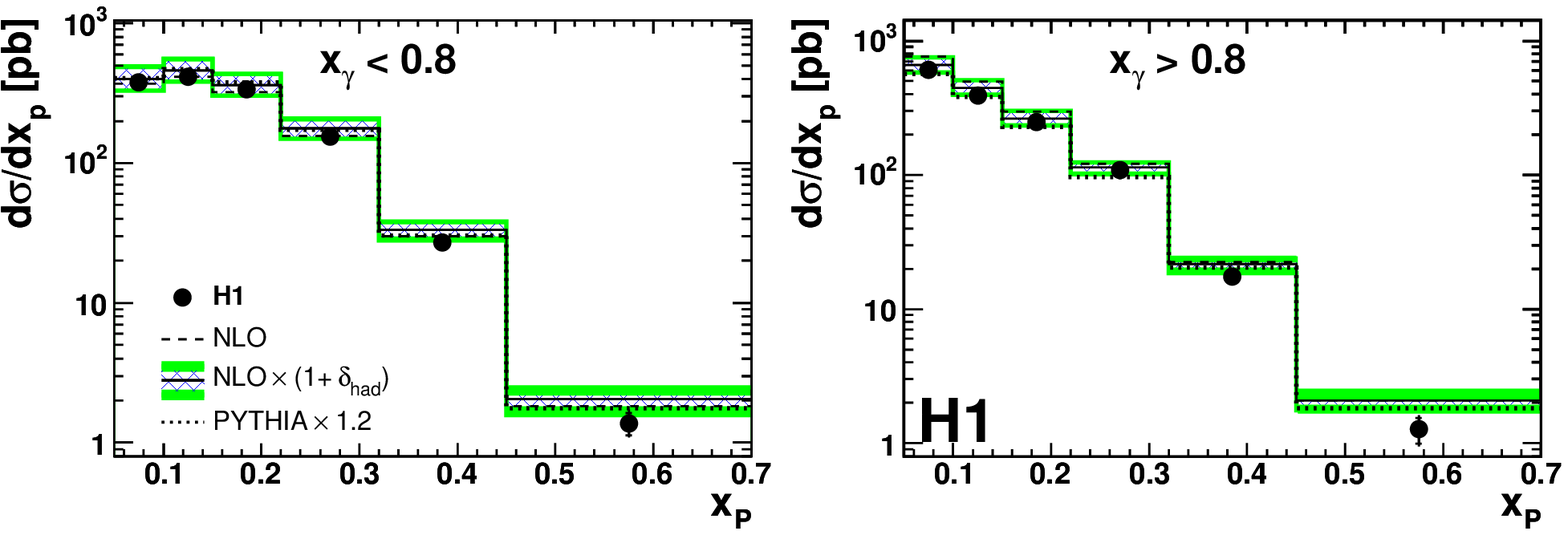}
    \caption{Bin averaged cross sections as a function of $x_p$ for data (points), NLO QCD predictions 
       with (solid line) and without (dashed line) hadronisation corrections $\delta_{\rm{had}}$ and for 
       the PYTHIA Monte Carlo predictions (dotted line) scaled by a factor of $1.2$. The inner bars indicate 
       the statistical uncertainty and the outer error bars show the statistical and systematic errors 
       added in quadrature. The inner (hatched) band of the NLO$\times (1+\delta_{\rm{had}})$ result is the 
       scale uncertainty, the outer (shaded) band is the total uncertainty. The cross sections are shown 
       separately for
       two regions in $x_\gamma$ enhancing the resolved (left) or direct (right) photon contribution.}
    \label{fig:xproton} 
  \end{center}
\end{figure}

The cross section as a function of $x_p$ is depicted in figure~\ref{fig:xproton}. Here the 
measurement is made in two regions of $x_\gamma$ ($x_\gamma > 0.8$ and $x_\gamma < 0.8$). In both 
regions the agreement of the NLO QCD predictions with the data is within $10\%$ at low $x_p$. 
This is covered by the experimental uncertainties which are dominated by the hadronic energy scale 
uncertainty. The two other significant contributions to the experimental uncertainty are the model 
uncertainty ($5\%$ at low $x_p$) and the statistical uncertainty ($\approx 20\%$ in the highest $x_p$ bin).

\begin{figure}[e]
  \begin{center}
    \includegraphics[width=\textwidth]{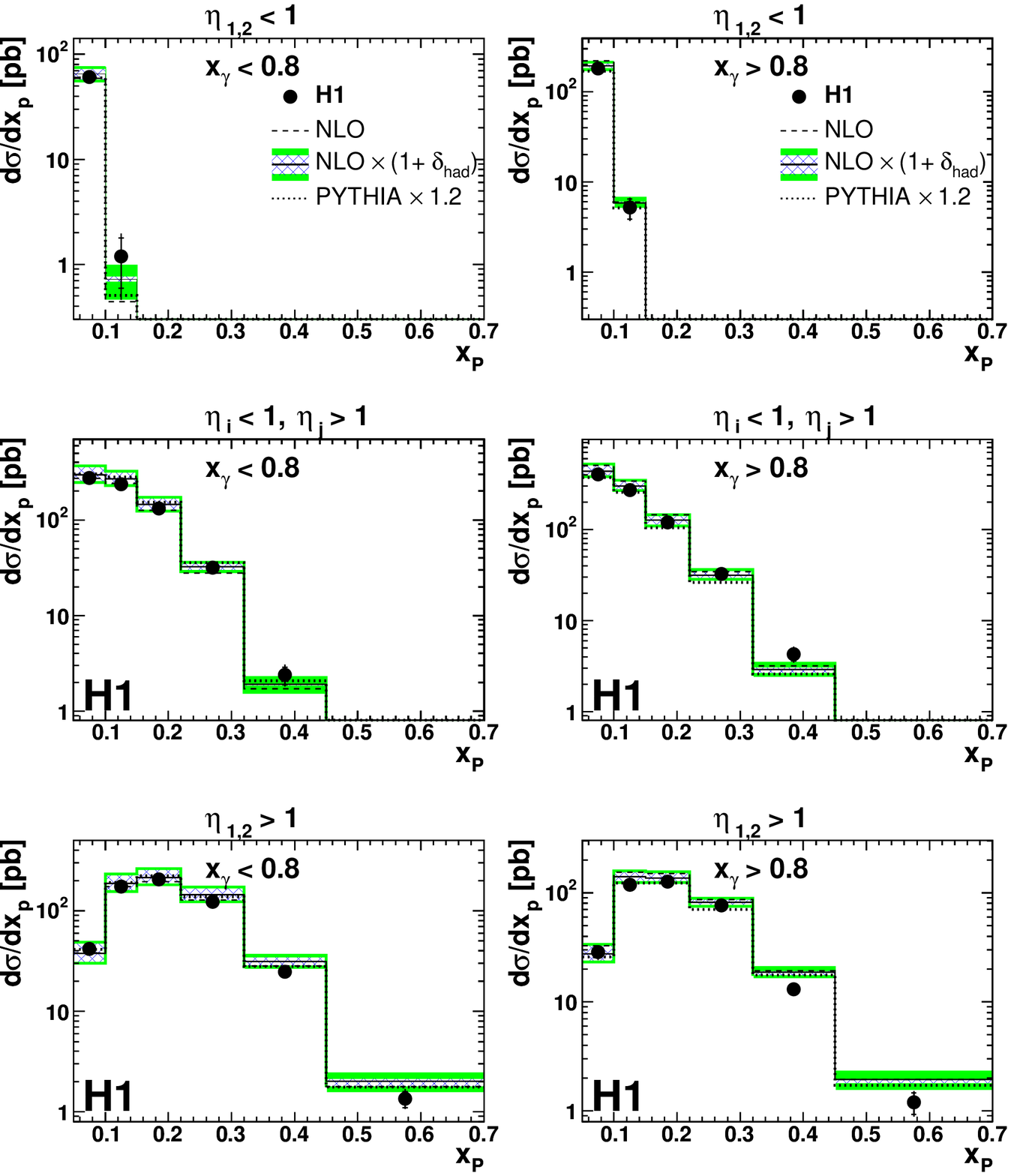}
    \caption{Bin averaged cross sections as a function of $x_p$ with different topologies of jet $\eta$ 
       for data (points), NLO QCD calculations with (solid line) and without (dashed line) hadronisation 
       corrections $\delta_{\rm{had}}$ and for the PYTHIA Monte Carlo predictions (dotted line) scaled by 
       a factor of $1.2$. The inner bars indicate the statistical uncertainty and the outer error bars show 
       the statistical and systematic errors added in quadrature. The inner (hatched) band of the 
       NLO$\times (1+\delta_{\rm{had}})$ result is the scale uncertainty, the outer (shaded) band is the total 
       uncertainty. The cross sections are shown separately for two regions in $x_\gamma$ enhancing the resolved (left) 
       or direct (right) photon contribution.}
    \label{fig:xpeta} 
  \end{center}
\end{figure}

\begin{figure}[e]
  \begin{center}
    \includegraphics[width=\textwidth]{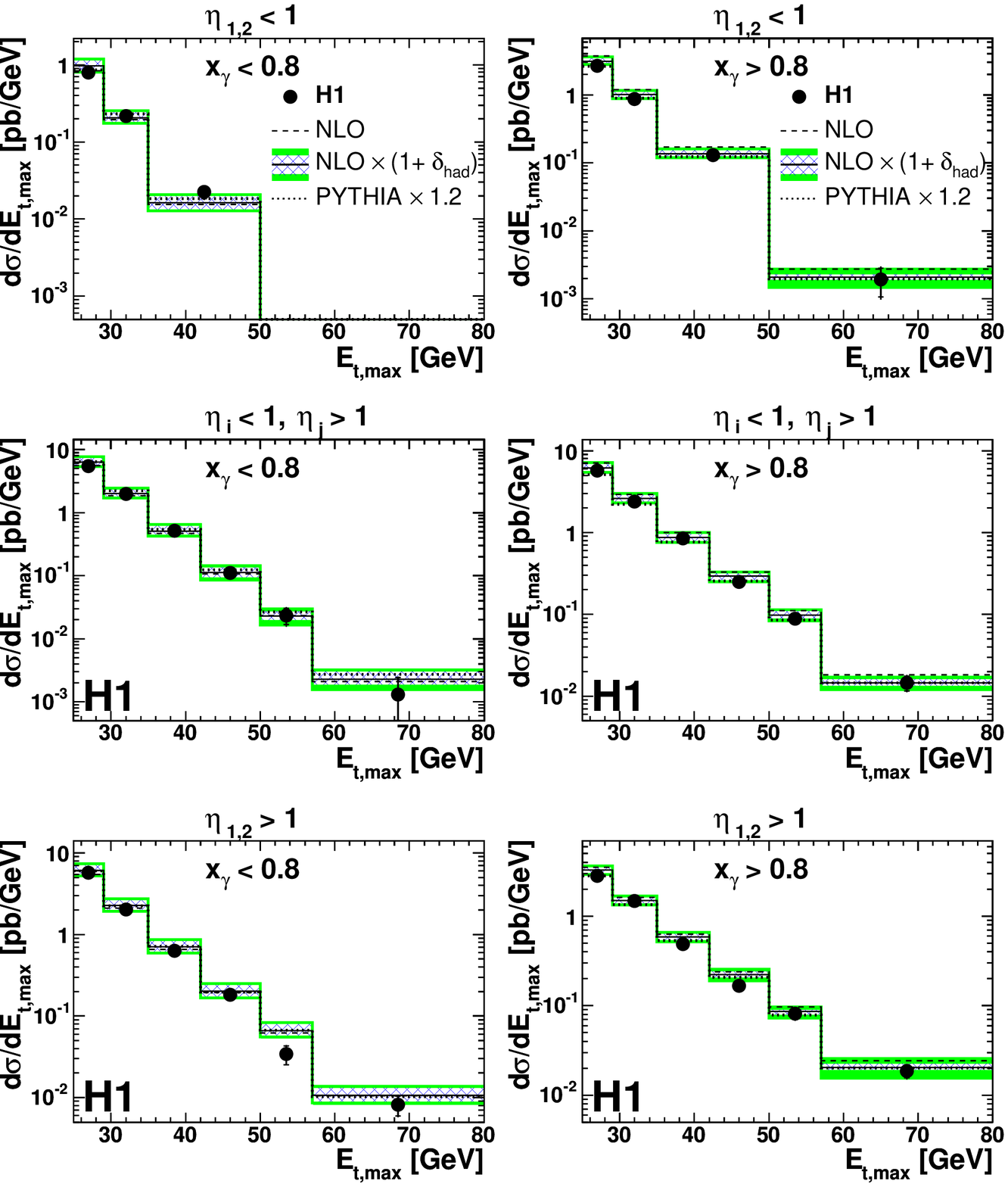}
    \caption{Bin averaged cross sections as a function of $E_{\rm{t,max}}$ with different topologies of 
       jet $\eta$ for data (points), NLO QCD calculations with (solid line) and without (dashed line) 
       hadronisation corrections $\delta_{\rm{had}}$ and for the PYTHIA Monte Carlo predictions (dotted line) 
       scaled by a factor of $1.2$. The inner bars indicate the statistical uncertainty and the outer error 
       bars show the statistical and systematic errors added in quadrature. The inner (hatched) band of 
       the NLO$\times (1+\delta_{\rm{had}})$ result is the scale uncertainty, the outer (shaded) band is the total 
       uncertainty. The cross sections are shown separately for two regions in $x_\gamma$ enhancing the resolved (left) or 
       direct (right) photon contribution.}
    \label{fig:ptmaxeta} 
  \end{center}
\end{figure}

Since the pseudorapidities of the two jets are sensitive to the momentum distributions of the interacting 
partons, the cross sections as a function of $x_p$ (figure~\ref{fig:xpeta}, table~\ref{table:xp}) and 
of $E_{\rm{t,max}}$ (figure~\ref{fig:ptmaxeta}, table~\ref{table:ptmax}) are measured for three different 
topologies of the final state: the case where both jets are in the ``backward" direction ($\eta_{1,2} < 1$), 
where both jets are in the ``forward" direction ($\eta_{1,2}>1$), and where one jet is in the 
``forward" ($\eta_{i}>1$) direction and one is in the ``backward" ($\eta_{j}<1$) direction. 
As before, the measurement is performed separately in two regions of $x_\gamma$. The NLO QCD predictions 
describe the data in $x_p$ and $E_{\rm{t,max}}$ well, except for large $x_p$ in the direct enhanced 
sample and with both jets going forward.

\section{Conclusion}
\noindent
In this paper a new and more precise measurement of high $E_{\rm{t}}$ dijet photoproduction is presented. 
Differential cross sections are measured in two regions of the observable $x_\gamma$. They are studied 
as a function of $|\cos\theta^{*}|$ and $x_p$. Furthermore the cross sections as a function of $E_{\rm{t,max}}$ 
and $x_p$ are investigated for different jet topologies. Both the NLO QCD calculation and the PYTHIA 
Monte Carlo simulation provide a reasonable description of the data.

The region of $x_\gamma > 0.8$ (direct photon enhanced region), in which the photon predominantly interacts 
directly with the proton, is particularly well suited to test proton structure as the photon structure 
plays no significant role there. At high $E_{\rm{t,max}}$ and large $x_{p}$ the dominant theoretical 
uncertainty comes from the uncertainty of the proton parton density functions. 
The data in the region of $x_\gamma < 0.8$ (resolved photon enhanced region), where the photon mainly 
behaves like a hadronic object, may also provide additional constraints on the 
photon parton density functions.

\section*{Acknowledgements}

We are grateful to the HERA machine group whose outstanding
efforts have made this experiment possible. 
We thank
the engineers and technicians for their work in constructing and
maintaining the H1 detector, our funding agencies for 
financial support, the
DESY technical staff for continual assistance
and the DESY directorate for support and for the
hospitality which they extend to the non DESY 
members of the collaboration.


\begin{table}[e] \small
  \centering
  \begin{tabular}{|c||c|c|c|c|c|}
    \hline
    $|\cos\theta^{*}|$ & $\frac{d\sigma}{d|\cos\theta^{*}|}$ [pb] & $\delta_{\rm{stat}}$ [\%] & $\delta_{\rm{tot}}$ [\%] &
    $\delta_{\rm{LAr}}$ [\%] & $\delta_{\rm{mod}}$ [\%] \\
    \hline 
    & \multicolumn{5}{c|}{$x_\gamma < 0.8$} \\ 
    \hline
0.00-0.10       & 104.9         &  3.4  & 10.7/8.1      & 9.6/5.0       & 1.5/2.8     \\
0.10-0.20       & 110.0         &  3.3  & 9.6/9.2       & 8.3/5.5       & 2.1/3.6     \\
0.20-0.30       & 105.2         &  3.5  & 11.1/7.7      & 9.4/5.7       & 3.1/3.1     \\
0.30-0.40       & 111.9         &  3.5  & 10.5/7.5      & 9.1/6.1       & 1.8/1.6     \\
0.40-0.50       & 111.0         &  3.5  & 10.8/6.8      & 9.5/5.2       & 1.8/1.6     \\
0.50-0.60       & 107.2         &  3.7  & 15.2/7.8      & 10.2/5.6      & 5.8/3.5     \\
0.60-0.70       &  91.65        &  3.9  & 14.0/7.0      & 10.2/4.6      & 4.8/3.0     \\
0.70-0.85       &  63.34        &  3.8  & 14.9/12.2     & 9.2/5.8       & 10.0/9.9    \\
    \hline 
    & \multicolumn{5}{c|}{$x_\gamma > 0.8$} \\ 
    \hline
0.00-0.10       & 121.6         &  3.1  & 7.5/9.0       & 6.3/5.0       & 1.3/3.4     \\
0.10-0.20       & 120.3         &  3.1  & 7.4/10.0      & 6.3/5.6       & 1.3/3.9     \\
0.20-0.30       & 124.3         &  3.1  & 8.0/5.7       & 6.9/3.5       & 1.3/1.7     \\
0.30-0.40       & 118.0         &  3.3  & 7.4/6.7       & 5.8/5.1       & 2.2/2.2     \\
0.40-0.50       & 114.2         &  3.4  & 8.5/6.0       & 7.2/4.2       & 1.4/1.4     \\
0.50-0.60       & 105.4         &  3.6  & 12.4/8.3      & 7.9/4.6       & 6.3/5.5     \\
0.60-0.70       &  95.70        &  3.8  & 9.1/6.8       & 6.7/4.7       & 2.9/2.3     \\
0.70-0.85       &  60.91        &  3.7  & 13.9/11.5     & 8.6/4.3       & 9.8/9.8     \\
    \hline 
    & \multicolumn{5}{c|}{$x_\gamma < 0.8$ and $M_{\rm{JJ}} > 65\gev$} \\ 
    \hline
0.00-0.10       &  15.19        &  9.4  & 14.7/12.3     & 9.8/5.1       & 3.6/4.1    \\
0.10-0.20       &  13.51        &  9.8  & 15.7/12.8     & 11.1/5.4      & 3.5/4.2    \\
0.20-0.30       &  14.38        & 10.1  & 16.3/11.2     & 11.1/3.1      & 3.3/3.3    \\
0.30-0.40       &  18.15        &  9.0  & 13.8/13.3     & 8.2/8.2       & 5.1/4.9    \\
0.40-0.50       &  23.61        &  7.8  & 13.3/10.3     & 9.2/5.8       & 3.0/2.7    \\
0.50-0.60       &  35.90        &  6.5  & 16.5/9.6      & 10.7/6.4      & 5.5/2.2    \\
0.60-0.70       &  56.38        &  5.1  & 15.1/6.8      & 10.9/3.4      & 4.5/2.2    \\
0.70-0.85       &  56.58        &  4.0  & 14.6/11.3     & 9.0/4.7       & 9.4/9.2    \\
    \hline 
    & \multicolumn{5}{c|}{$x_\gamma > 0.8$ and $M_{\rm{JJ}} > 65\gev$} \\ 
    \hline
0.00-0.10       &  30.99        &  6.7  & 10.4/11.0     & 7.1/6.0       & 2.6/3.8     \\
0.10-0.20       &  29.54        &  6.9  & 12.2/12.1     & 8.4/5.4       & 5.1/6.0     \\
0.20-0.30       &  33.16        &  6.7  & 10.3/8.8      & 7.1/4.5       & 2.5/2.6     \\
0.30-0.40       &  35.60        &  6.6  & 10.5/8.7      & 7.1/4.8       & 2.4/2.3     \\
0.40-0.50       &  40.79        &  6.2  & 10.4/8.2      & 7.6/4.4       & 2.2/2.2     \\
0.50-0.60       &  53.04        &  5.5  & 12.4/7.6      & 7.8/4.4       & 4.0/2.0     \\
0.60-0.70       &  74.15        &  4.6  & 9.4/7.0       & 6.8/4.4       & 2.7/1.9     \\
0.70-0.85       &  58.86        &  3.9  & 12.8/10.1     & 8.6/4.1       & 8.1/8.1     \\
    \hline
  \end{tabular}
  \caption{Bin averaged cross sections for dijet photoproduction in intervals of
     $|\cos\theta^{*}|$ shown with the statistical error ($\delta_{\rm{stat}}$), the total 
     error including statistical and systematic errors ($\delta_{\rm{tot}}$), the error coming 
     from the LAr hadronic energy scale uncertainty ($\delta_{\rm{LAr}}$) 
     and the error from the model 
     uncertainty and the Monte Carlo reweighting ($\delta_{\rm{mod}}$). 
     Two numbers are shown to allow for asymmetric errors ($+/-$).}
    \label{table:costheta} 
\end{table}

\begin{table}[htbp] \small
  \centering
  \begin{tabular}{|c||c|c|c|c|c|}
    \hline
    $x_\gamma$ & $\frac{d\sigma}{dx_\gamma}$ [pb] & $\delta_{\rm{stat}}$ [\%] & $\delta_{\rm{tot}}$ [\%] &
    $\delta_{\rm{LAr}}$ [\%] & $\delta_{\rm{mod}}$ [\%] \\
    \hline 
    & \multicolumn{5}{c|}{$x_p < 0.1$} \\ 
    \hline
0.30-0.50       &  19.11        &  5.8  & 11.9/10.4     & 7.4/7.0       & 4.6/3.4     \\
0.50-0.70       &  46.43        &  3.5  & 12.0/10.3     & 7.1/5.9       & 7.6/7.2     \\
0.70-0.85       &  94.58        &  2.9  & 10.3/9.1      & 6.9/5.7       & 6.6/6.2     \\
0.85-1.00       & 220.7         &  1.8  & 8.4/8.5       & 5.4/5.5       & 5.8/5.8     \\
    \hline 
    & \multicolumn{5}{c|}{$x_p > 0.1$} \\ 
    \hline
0.10-0.30       &  55.51        &  4.0  & 14.1/8.0      & 12.3/6.3      & 2.4/1.7     \\
0.30-0.50       &  90.88        &  2.8  & 12.4/6.4      & 10.9/4.8      & 2.7/2.6     \\
0.50-0.70       & 103.8         &  2.5  & 10.9/6.4      & 9.6/5.0       & 2.2/2.2     \\
0.70-0.85       & 148.7         &  2.4  & 9.9/7.2       & 8.9/5.5       & 2.4/2.7     \\
0.85-1.00       & 276.5         &  1.9  & 8.6/6.0       & 7.8/3.5       & 0.7/2.1     \\
    \hline
  \end{tabular}
   \caption{Bin averaged cross sections for dijet photoproduction in intervals of $x_\gamma$ 
       shown with the statistical error ($\delta_{\rm{stat}}$), the total error including statistical 
       and systematic errors ($\delta_{\rm{tot}}$), the error coming from the LAr hadronic energy scale 
       uncertainty ($\delta_{\rm{LAr}}$) and the error from the model uncertainty and the Monte Carlo 
       reweighting ($\delta_{\rm{mod}}$). 
       Two numbers are shown to allow for asymmetric errors ($+/-$).
       }
    \label{table:xgamma} 
\end{table}

\begin{table}[htbp] \small
  \centering
  \begin{tabular}{|c||c|c|c|c|c|c|}
    \hline
    $x_p$ & $\frac{d\sigma}{dx_p}$ [pb] & $\delta_{\rm{stat}}$ [\%] & $\delta_{\rm{tot}}$ [\%] &
    $\delta_{\rm{LAr}}$ [\%]  & $\delta_{\rm{mod}}$ [\%] \\
    \hline 
    & \multicolumn{5}{c|}{$x_\gamma < 0.8$ and $\eta_{1,2} < 1$} \\ 
    \hline
0.05-0.10       &  60.60        &  6.3  & 11.9/11.8     & 7.2/7.2       & 6.1/5.9    \\
0.10-0.15       &   1.18        & 50.0  & 65.2/63.9     & 30.0/30.0     & 26.7/25.8  \\
    \hline 
    & \multicolumn{5}{c|}{$x_\gamma < 0.8$ and $\eta_{i} < 1, \eta_{j} > 1$} \\ 
    \hline
0.05-0.10       & 276.2         &  3.0  & 10.9/9.0      & 6.7/6.1       & 6.1/5.4     \\
0.10-0.15       & 235.4         &  3.6  & 13.1/8.0      & 9.3/5.2       & 5.5/4.5     \\
0.15-0.22       & 132.0         &  4.0  & 13.3/8.1      & 10.3/5.7      & 4.4/3.6     \\
0.22-0.32       &  31.80        &  6.4  & 16.4/12.0     & 10.5/4.2      & 9.2/9.0     \\
0.32-0.45       &   2.38        & 20.5  & 28.1/24.3     & 14.1/7.7      & 10.9/10.3   \\
    \hline 
    & \multicolumn{5}{c|}{$x_\gamma < 0.8$ and $\eta_{1,2} > 1$} \\ 
    \hline
0.05-0.10       &  41.84        &  6.8  & 11.4/11.8     & 8.1/7.1       & 3.7/4.4     \\
0.10-0.15       & 175.1         &  3.9  & 11.2/8.5      & 9.5/5.3       & 3.6/4.0     \\
0.15-0.22       & 205.5         &  3.1  & 10.5/7.3      & 9.3/5.8       & 1.4/1.7     \\
0.22-0.32       & 122.7         &  3.3  & 13.7/6.6      & 12.4/4.4      & 1.4/2.0     \\
0.32-0.45       &  24.69        &  6.1  & 14.0/10.1     & 11.8/5.5      & 2.3/3.4     \\
0.45-0.70       &   1.35        & 18.4  & 25.5/21.1     & 15.2/5.2      & 6.4/7.0     \\
    \hline 
    & \multicolumn{5}{c|}{$x_\gamma > 0.8$ and $\eta_{1,2} < 1$} \\ 
    \hline
0.05-0.10       & 180.5         &  3.8  & 8.1/7.8       & 6.2/5.8       & 2.9/2.9     \\
0.10-0.15       &   5.17        & 25.0  & 29.4/29.2     & 12.0/12.0     & 9.0/8.8     \\
    \hline 
    & \multicolumn{5}{c|}{$x_\gamma > 0.8$ and $\eta_{i} < 1, \eta_{j} > 1$} \\ 
    \hline
0.05-0.10       & 401.9         &  2.4  & 8.7/8.4       & 5.6/5.2       & 5.8/5.8     \\
0.10-0.15       & 269.6         &  3.3  & 9.0/6.6       & 7.0/4.0       & 3.6/3.6     \\
0.15-0.22       & 120.1         &  4.0  & 9.9/7.3       & 7.8/4.5       & 3.6/3.6     \\
0.22-0.32       &  32.43        &  6.4  & 12.3/8.5      & 9.4/4.2       & 2.8/2.8     \\
0.32-0.45       &   4.27        & 16.0  & 21.3/19.8     & 10.8/8.2      & 8.0/7.9     \\
    \hline 
    & \multicolumn{5}{c|}{$x_\gamma > 0.8$ and $\eta_{1,2} > 1$} \\ 
    \hline
0.05-0.10       &  28.58        &  7.7  & 11.2/15.5     & 6.2/5.5       & 4.5/7.2    \\
0.10-0.15       & 118.6         &  4.5  & 7.4/11.8      & 5.2/4.5       & 1.8/5.1    \\
0.15-0.22       & 126.7         &  4.1  & 10.7/9.8      & 8.8/2.7       & 2.1/4.5    \\
0.22-0.32       &  76.27        &  4.6  & 11.9/10.0     & 9.7/3.2       & 3.5/5.0    \\
0.32-0.45       &  13.00        &  9.3  & 14.5/12.9     & 9.5/3.7       & 3.1/4.8    \\
0.45-0.70       &   1.20        & 22.4  & 27.4/26.2     & 13.8/8.8      & 7.1/8.0    \\
    \hline
  \end{tabular}
    \caption{Bin averaged cross sections for dijet photoproduction in intervals of $x_p$ shown 
        with the statistical error ($\delta_{\rm{stat}}$), the total error including statistical 
        and systematic errors ($\delta_{\rm{tot}}$), the error coming from the LAr hadronic energy scale 
        uncertainty ($\delta_{\rm{LAr}}$) and the error from the model uncertainty and the Monte Carlo 
        reweighting ($\delta_{\rm{mod}}$). 
        Two numbers are shown to allow for asymmetric errors ($+/-$).}
    \label{table:xp}
\end{table}

\begin{table}[htbp] \small
  \centering
  \begin{tabular}{|c||c|c|c|c|c|}
    \hline
    $E_{\rm{t,max}}$ [\gev] & $\frac{d\sigma}{dE_{\rm{t,max}}}$ $[\frac{\rm pb} {\gev}]$ & $\delta_{\rm{stat}}$ [\%] & $\delta_{\rm{tot}}$ [\%] &
    $\delta_{\rm{LAr}}$ [\%] & $\delta_{\rm{mod}}$ [\%] \\
    \hline 
    & \multicolumn{5}{c|}{$x_\gamma < 0.8$ and $\eta_{1,2} < 1$} \\ 
    \hline
 25-29  &  0.796        &  6.1  & 16.5/14.3     & 8.5/4.1       & 12.0/11.8   \\
 29-35  &  0.217        &  9.2  & 12.5/13.7     & 4.5/8.8       & 4.9/4.0     \\
 35-50  &  0.022        & 17.8  & 22.7/21.1     & 7.4/7.8       & 8.9/7.7     \\
    \hline 
    & \multicolumn{5}{c|}{$x_\gamma < 0.8$ and $\eta_{i} < 1, \eta_{j} > 1$} \\ 
    \hline
 25-29  &  5.500        &  2.6  & 11.0/8.0      & 7.8/5.4       & 5.4/4.9     \\
 29-35  &  1.989        &  3.3  & 13.1/8.3      & 9.9/5.8       & 5.3/4.4     \\
 35-42  &  0.517        &  5.8  & 15.4/10.7     & 9.0/5.6       & 7.8/6.7     \\
 42-50  &  0.110        & 12.0  & 20.3/14.9     & 10.5/5.2      & 8.3/6.8     \\
 50-57  &  0.023        & 28.9  & 36.5/34.0     & 15.0/15.0     & 11.6/9.6    \\
 57-80  &  0.001        & 86.4  & 92.4/90.4     & 20.0/20.0      & 19.1/17.0   \\
    \hline 
    & \multicolumn{5}{c|}{$x_\gamma < 0.8$ and $\eta_{1,2} > 1$} \\ 
    \hline
 25-29  &  5.691        &  2.6  & 11.6/6.6      & 10.7/4.4      & 1.2/2.1     \\
 29-35  &  2.031        &  3.0  & 11.1/8.0      & 9.9/6.2       & 1.2/2.0     \\
 35-42  &  0.631        &  5.2  & 12.4/8.3      & 10.7/5.4      & 2.0/2.3     \\
 42-50  &  0.181        &  8.9  & 15.6/12.5     & 11.2/7.8      & 3.3/3.4     \\
 50-57  &  0.034        & 25.7  & 29.5/27.7     & 12.5/7.0      & 6.6/6.6     \\
 57-80  &  0.008        & 27.0  & 32.7/29.3     & 14.8/7.0      & 8.6/8.4     \\
    \hline 
    & \multicolumn{5}{c|}{$x_\gamma > 0.8$ and $\eta_{1,2} < 1$} \\ 
    \hline
 25-29  &  2.664        &  3.3  & 8.8/8.8       & 5.8/5.8       & 5.3/5.3     \\
 29-35  &  0.872        &  4.6  & 9.6/9.3       & 5.5/5.2       & 5.9/5.8     \\
 35-50  &  0.131        &  7.5  & 12.0/11.2     & 6.3/6.5       & 5.3/4.8     \\
 50-80  &  0.002        & 44.7  & 54.2/52.3     & 21.0/21.0     & 18.3/17.0   \\
    \hline 
    & \multicolumn{5}{c|}{$x_\gamma > 0.8$ and $\eta_{i} < 1, \eta_{j} > 1$} \\ 
    \hline
 25-29  &  5.751        &  2.4  & 9.0/7.6       & 6.4/4.5       & 5.3/5.3     \\
 29-35  &  2.400        &  2.9  & 8.6/7.4       & 6.5/5.1       & 4.1/4.1     \\
 35-42  &  0.854        &  4.6  & 10.2/7.6      & 7.5/4.5       & 3.8/3.5     \\
 42-50  &  0.249        &  8.2  & 13.4/10.4     & 8.2/5.0       & 4.4/3.5     \\
 50-57  &  0.089        & 14.5  & 19.1/16.6     & 7.9/5.7       & 6.5/5.4     \\
 57-80  &  0.015        & 20.3  & 25.6/23.9     & 10.0/10.0     & 8.5/7.1     \\
    \hline 
    & \multicolumn{5}{c|}{$x_\gamma > 0.8$ and $\eta_{1,2} > 1$} \\ 
    \hline
 25-29  &  2.833        &  3.7  & 7.9/10.7      & 6.3/3.5       & 1.8/4.8     \\
 29-35  &  1.474        &  3.9  & 11.0/10.4     & 9.4/2.5       & 2.1/4.9     \\
 35-42  &  0.487        &  6.0  & 11.7/12.4     & 8.3/5.1       & 4.5/6.1     \\
 42-50  &  0.167        &  9.6  & 14.0/13.7     & 8.7/5.4       & 3.9/5.2     \\
 50-57  &  0.081        & 14.0  & 18.3/16.7     & 10.2/4.7      & 5.7/6.2     \\
 57-80  &  0.019        & 16.7  & 19.8/19.7     & 8.1/7.6       & 6.2/6.4     \\
    \hline
  \end{tabular}
   \caption{Bin averaged cross sections for dijet photoproduction in intervals of $E_{\rm{t,max}}$ 
      shown with the statistical error ($\delta_{\rm{stat}}$), the total error including statistical 
      and systematic errors ($\delta_{\rm{tot}}$), the error coming from the LAr hadronic energy scale 
      uncertainty ($\delta_{\rm{LAr}}$) and the error from the model uncertainty and the Monte Carlo 
      reweighting ($\delta_{\rm{mod}}$). 
      Two numbers are shown to allow for asymmetric errors ($+/-$).}
    \label{table:ptmax} 
\end{table}

\end{document}